\documentclass[reprint,aps,pra,superscriptaddress]{revtex4-1}

\usepackage{graphicx}
\usepackage{amsmath}
\usepackage{amssymb}
\usepackage{dcolumn}
\usepackage{bm}
\usepackage{setspace}
\usepackage{color}
\usepackage[capitalize,nameinlink]{cleveref}

\begin{document}

\title{\LARGE Conformational dynamics and phase behavior of lipid vesicles in a precisely controlled extensional flow}

\author{Dinesh Kumar}
\affiliation
{
	Department of Chemical and Biomolecular Engineering \\ University of Illinois at Urbana-Champaign, Urbana, IL, 61801
}
\affiliation
{
	Beckman Institute for Advanced Science and Technology \\ University of Illinois at Urbana-Champaign, Urbana, IL, 61801
}

\author{Channing M. Richter}
\affiliation
{
	Department of Chemical and Biomolecular Engineering \\ University of Illinois at Urbana-Champaign, Urbana, IL, 61801
}

\author{Charles M. Schroeder}
\email[To whom correspondence must be addressed: ]{cms@illinois.edu}
\affiliation
{
	Department of Chemical and Biomolecular Engineering \\ University of Illinois at Urbana-Champaign, Urbana, IL, 61801
}
\affiliation
{
	Beckman Institute for Advanced Science and Technology \\ University of Illinois at Urbana-Champaign, Urbana, IL, 61801
}
\affiliation
{
	Department of Materials Science and Engineering \\ University of Illinois at Urbana-Champaign, Urbana, IL, 61801
}

\date{\today}

\begin{abstract}
	Lipid vesicles play a key role in fundamental biological processes. Despite recent progress, we lack a complete understanding of the non-equilibrium dynamics of vesicles due to challenges associated with long-time observation of shape fluctuations in strong flows. In this work, we present a flow-phase diagram for vesicle shape and conformational transitions in planar extensional flow using a Stokes trap, which enables control over the center-of-mass position of single or multiple vesicles in precisely defined flows [Shenoy, Rao, Schroeder, \textit{PNAS}, 113(15):3976-3981, 2016]. In this way, we directly observe the non-equilibrium conformations of lipid vesicles as a function of reduced volume $\nu$, capillary number $Ca$, and viscosity contrast $\lambda$. Our results show that vesicle dynamics in extensional flow are characterized by the emergence of three distinct shape transitions, including a tubular to symmetric dumbbell transition, a spheroid to asymmetric dumbbell transition, and quasi-spherical to ellipsoid transition. The experimental phase diagram is in good agreement with recent predictions from simulations [Narsimhan, Spann, Shaqfeh, \textit{J. Fluid Mech.}, 2014, \textbf{750}, 144]. We further show that the phase boundary of vesicle shape transitions is independent of the viscosity contrast. Taken together, our results demonstrate the utility of the Stokes trap for the precise quantification of vesicle stretching dynamics in precisely defined flows.
\end{abstract}

\maketitle

\section{Introduction}
Vesicles are fluid-filled soft containers enclosed by a molecularly thin (3-4 nm) lipid bilayer membrane suspended in a liquid medium. In recent years, the mechanics of giant unilamellar vesicles (GUVs) has been extensively studied to provide insight into the mechanical properties of biological systems such as red blood cells \cite{boal2002mechanics,fenz2012giant}. To this end, vesicles have been used to understand the equilibrium and non-equilibrium dynamics of simplified cells that do not contain a cytoskeleton or a polymerized, protein-laden membrane commonly found in living cells \cite{abreu2014fluid,freund2014numerical}. Artificial vesicles have also been used for the triggered release of cargo in biomedical applications such as drug delivery and micro/nanoscale reactors \cite{weiner1989liposomes,honeywell2005vesicles,uchegbu1998non}.

Achieving a full understanding of the non-equilibrium dynamics of single-component lipid vesicles in precisely defined flows is crucial for understanding cell mechanics. From this view, such studies can inform how the fluid dynamics and membrane properties inside and outside the fluid-filled compartment contribute to cell shape changes. From this view, a significant amount of prior work has been focused on investigating the shape dynamics of vesicles under different flow conditions, such as Poiseuille flow \cite{danker2009vesicles,noguchi2005shape,coupier2008noninertial}, shear flow \cite{kantsler2005orientation,deschamps2009phase,hatakenaka2011orientation,honerkamp2013membrane,de1997deformation,shahidzadeh1998large,deschamps2009dynamics,kantsler2006transition,mader2006dynamics,vlahovska2007dynamics,misbah2006vacillating,biben2011three,zhao2011dynamics,zabusky2011dynamics,kantsler2008dynamics}, and extensional flow \cite{kantsler2008critical,kantsler2007vesicle,lebedev2007dynamics,zhao2013shape,narsimhan2015pearling,narsimhan2014mechanism,dahl2016experimental}. Experiments and simulations on vesicles in shear flow have uncovered intriguing dynamic behavior including: (i) tumbling, where a vesicle undergoes a periodic flipping motion, (ii) trembling, where vesicle shape fluctuates and the orientation oscillates in time, and (iii) tank-treading, where an ellipsoid vesicle's major axis maintains a fixed orientation with respect to the flow direction while the membrane rotates about the vorticity axis \cite{kantsler2005orientation,kantsler2006transition,zhao2011dynamics,misbah2006vacillating}. The transitions between these dynamical motions depend on shear rate $\dot{\gamma}$, viscosity ratio $\lambda$ between the inner $\mu_{in}$ and outer $\mu_{out}$ fluid viscosities, and reduced volume $\nu$, which is a measure of vesicle's asphericity \cite{misbah2012vesicles}. Prior work has also focused on the induced hydrodynamic lift of a single vesicle near a wall in shear flow \cite{callens2008hydrodynamic}, pair interactions between vesicles in flow \cite{gires2012hydrodynamic,gires2014pairwise}, and measurement of the effective viscosity of a dilute vesicle suspension \cite{honerkamp2013membrane}.

Despite recent progress in understanding vesicle dynamics in shear flow, the behavior of vesicles in extensional flow is less well understood. Extensional flow is considered to be a strong flow that can induce high levels of membrane deformation. Unlike simple shear, extensional flows consist of purely extensional-compressional character without elements of fluid rotation \cite{leal2007advanced}. In natural blood flows, red blood cells repeatedly undergo reversible deformations by a combination of shear and extension when passing through capillaries in the body \cite{braunmuller2012hydrodynamic,lee2009extensional}.   From this perspective, there is a clear need to understand the shape dynamics of cells when transiting through narrow capillaries. Interestingly, prior work has shown that the extensional components of the velocity gradient tensor are crucial for predicting rupture of red blood cells undergoing tank-treading motion in shear flow \cite{down2011significance,lee2009extensional}. From kinematic analysis, any general linear flow can be decomposed into elements of rotation and extension/compression, thereby identifying the flow field components associated with rupture. From this view, understanding the dynamics of vesicles in extensional flow is of fundamental interest to elucidate dynamics in more complex mixed flows containing arbitrary amounts of rotation and extension/compression \cite{batchelor2000introduction,deschamps2009dynamics}. 

The physical properties of vesicles govern their dynamic behavior in flow. Reduced volume $\nu$ is defined as the ratio of a vesicle's volume $V$ to the volume of a sphere with an equivalent surface area $A$, such that $\nu=3V/4\pi R^3$ where $R=\sqrt{A/4\pi}$ is the vesicle's equivalent radius based on the total surface area. For a vesicle with a perfectly spherical shape, the reduced volume $\nu=1$, which means that there is no excess area to deform when subjected to hydrodynamic stress. In the weak-flow limit, a vesicle is described by a constant total surface area \cite{seifert1997configurations} and substantially deforms only if the reduced volume $\nu<1$. 

In 2008, Steinberg and coworkers studied the dynamics of highly deflated vesicles $\nu< 0.56$ in extensional flow \cite{kantsler2008critical}. At low reduced volumes, vesicles essentially adopt tubular shapes at equilibrium and are highly deformable due to the large surface area to volume ratio. Under these conditions, the dynamics of highly deflated vesicles in extensional flow was observed to be similar to the coil-stretch transition for flexible polymers in extensional flow \cite{larson1999structure, perkins1999single, schroeder2018single}. Above a critical strain rate $\dot{\epsilon_c}$, deflated vesicles were found to undergo a shape transition from a tubular to a symmetric dumbbell conformation. Steinberg and coworkers reported a flow phase-stability diagram for such shape transitions \cite{kantsler2008critical}, however, they did not directly characterize the bending modulus $\kappa_b$ for the deflated vesicles, and rather used an order-of-magnitude estimate of $\kappa_b$ from literature. Nevertheless, these experimental observations are in good agreement with numerical simulations by Shaqfeh and coworkers \cite{narsimhan2014mechanism, narsimhan2015pearling}, which further confirm the tubular-to-symmetric dumbbell shape transition for deflated vesicles. Moreover, these simulations examined vesicle dynamics under a wide range of reduced volumes, predicting that moderately deflated vesicles ($0.56<\nu<0.75$) would undergo a spheroid to asymmetric dumbbell shape transition due to destabilizing curvature changes in the membrane as a result of modified Rayleigh-Plateau mechanism \cite{narsimhan2014mechanism, narsimhan2015pearling}.

Recently, Muller and coworkers \cite{dahl2016experimental} studied the dynamics of lipid vesicles in planar extensional flow using a cross-slot microfluidic device \cite{schroeder2003observation}. The results from this study generally confirmed the spheroid to asymmetric dumbbell shape transition for moderately deflated ($0.56<\nu<0.75$) vesicles in extensional flow. Interestingly, Muller and coworkers reported a stability boundary for shape transitions in reduced volume-capillary number $(\nu, Ca)$ phase-space, where the capillary number $Ca=\mu_{out}\dot{\epsilon}R^3/\kappa_b$ is the ratio of the bending time scale to the flow time scale, $\dot{\epsilon}$ is the fluid strain rate, and $\mu_{out}$ is the viscosity of suspending medium. These experiments generally involved manual trapping of single vesicles near the stagnation point of planar extensional flow, which makes it challenging to observe dynamics over long times while maintaining a stable center-of-mass position of vesicles in flow. From this perspective, studying the non-equilibrium shape dynamics of lipid vesicles in precisely defined extensional flows is critically needed to understand vesicle shape transitions and stability in strong flows.

In this paper, we present a detailed flow-phase diagram of non-equilibrium vesicle shape transitions in extensional flow using a Stokes trap \cite{Shenoy12042016,kumar2019orientation,shenoy2019flow}, which enables precise control over the center-of-mass position of single or multiple particles in flow. In this way, we directly observe vesicle shape transitions over long observation times across a wide range of parameters including reduced volume $\nu$ and capillary number $Ca$. We first discuss the implementation of the Stokes trap technique in a PDMS-based microfluidic device. We then present a method to estimate the bending modulus of a vesicle at equilibrium using thermal fluctuation analysis. Using this approach, we systematically determine the flow-phase diagram for vesicles in $(\nu, Ca)$ space, and we investigate the effect of viscosity contrast $\lambda$ on the phase boundary. Finally, we discuss how the Stokes trap technique can be used to investigate the transient stretching and relaxation dynamics of vesicles under highly non-equilibrium flow conditions. 

\section{Experimental Methods}
\subsection{GUV preparation}
Giant unilamellar vesicles (GUVs) are prepared from a mixture of 1,2-dioleoyl-sn-glycero-3-phosphocholine (DOPC, Avanti Polar Lipids) and 0.12 mol \% of 1,2-dioleoyl-sn-glycero-3-phosphoethanolamine-N-(lissamine rhodamine B sulfonyl) (DOPE-Rh, Avanti Polar Lipids) using the classical electroformation method described in Angelova \textit{et al.} \cite{angelova1992preparation}. The fluorescently labeled lipid DOPE-Rh contains a rhodamine dye (absorption/emission maxima 560 nm/583 nm) on the lipid head group, rather than the tail group, because it is known that lipids with labeled hydrocarbon tails can result in altered membrane properties if the charged dye molecule flips into the hydrophilic head group space, which may affect the bending modulus of the membrane \cite{raghuraman2007monitoring}.

For electroformation of GUVs, a stock lipid solution is prepared with 25 mg/mL DOPC and 0.04 mg/mL DOPE-Rh. Next, 10 $\mu$L of the lipid solution in chloroform is spread on a conductive indium tin oxide (ITO) coated glass slide (resistance $\Omega$, 25$\times$50 $\times$1.1 mm, Delta Technologies) and dried under vacuum overnight. The pair of ITO slides are sandwiched together using a 1.5 mm Teflon spacer, forming a chamber with a volume of $\approx$2.4 mL and coupled to a function generator (Agilent 33220 A). The electroformation chamber is filled with 100 mM sucrose solution (Sigma-Aldrich) and an alternating current (AC) electric field of 2 V/mm at 10 Hz is applied for 120 min at room temperature (22$^\circ$C). Under these conditions, DOPC lipid remains in the fluid phase \cite{kantsler2008critical}. Most of the vesicles prepared by this method are unilamellar with few defects in the size range of 5-25 $\mu m$ in radius. The viscosity of the 100 mM sucrose solution ($\mu=1.1$ mPa-s) is measured using a benchtop viscometer (Brookfield) at 22$^\circ$C. Following electroformation, most vesicles are only weakly deflated and quasi-spherical in nature. To generate moderately deflated (low reduced volume) vesicles, 100 $\mu$L of a 200 mM sucrose solution is added to 2.0 mL of the electroformed vesicle suspension, which increases the total sucrose concentration to 105 mM. In this way, osmotic pressure differences tend to drive water out of the vesicle interior until the sucrose concentrations are nearly equal on both sides of the membrane \cite{dahl2016experimental,dimova2014recent}. The osmotic deflation method generated reduced volume vesicles in the range $0.30<\nu<0.90$, though the occurrence of extremely low reduced volume vesicles ($0.30<\nu<0.50$) in the suspension was relatively rare. For experiments involving high solvent viscosities (viscosity ratio $\lambda$ = 0.1), the viscosity of the suspending medium was increased to $\mu_{out}$ = 10.4 mPa-s by adding glycerol to the 100 mM sucrose solution.

\subsection{Stokes trap}
\begin{figure}[t]
	\centering
	\includegraphics[width=0.45\textwidth]{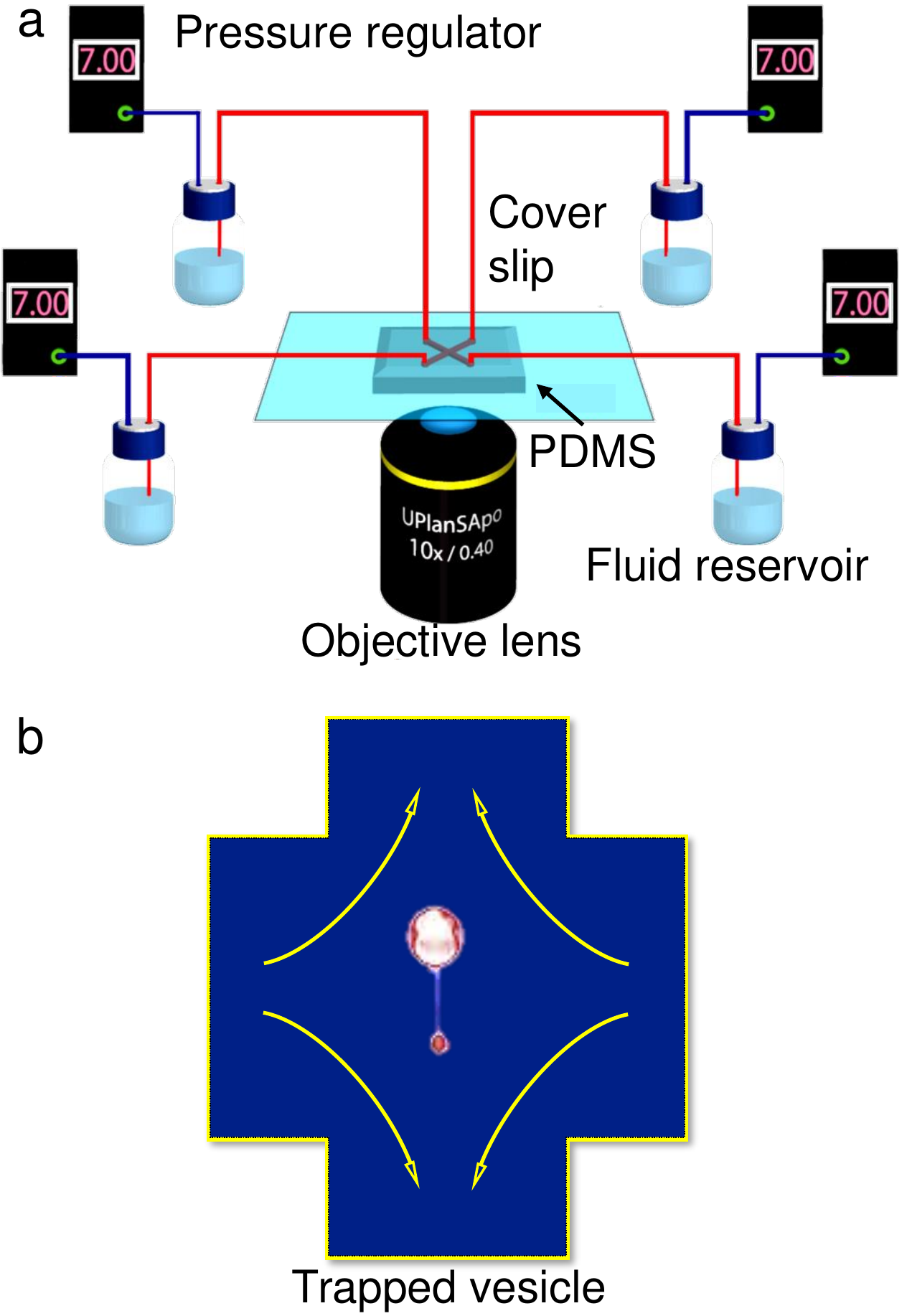}
	\caption{\label{vesicle_fig1}Stokes trap for studying vesicle dynamics in flow. (a) Schematic of the experimental setup used to generate planar extensional flow. Inlet/outlet channels in the microfluidic device are connected to fluidic reservoirs containing the vesicle suspension and pressurized by regulators controlled by a custom LabVIEW program, thereby generating pressure-driven flow in the cross-slot. (b) Schematic of the cross-slot microfluidic device showing a deformed vesicle trapped in extensional flow near the stagnation point for illustrative purposes (not drawn to scale). The width of channels is 400 $\mu$m and the radii of the two spherical ends of the deformed vesicle are 12 $\mu$m and 4 $\mu$m, respectively.}
\end{figure}

We use a Stokes trap \cite{Shenoy12042016} to generate controlled strain rate schedules while simultaneously achieving long-term confinement of single vesicles near the stagnation point of a planar extensional flow. A four-channel cross-slot microfluidic device is used for studying vesicle dynamics (Fig. \ref{vesicle_fig1}a). In brief, single-layer polydimethylsiloxane (PDMS)-based microfluidic device (width = 400 $\mu$m, and depth = 100 $\mu$m) is fabricated using standard techniques in soft lithography \cite{xia1998soft}. The channel dimensions are much larger compared to the typical vesicle equilibrium size $R$ = 5-25 $\mu$m, such that the effect of confinement is negligible. During device operation, fluid is injected into two opposing inlet channels and withdrawn through the two remaining outlet channels, thereby forming mutually perpendicular inlets and outlets. In this way, the symmetry of the flow-field under low Reynolds number conditions results in the formation of a fluid stagnation point (zero-velocity point) near the center of the cross-slot device, thereby generating a planar extensional flow in the vicinity of stagnation point as shown in Fig. \ref{vesicle_fig1}b. 

The Stokes trap was used to enable the direct observation of vesicle dynamics in extensional flow with a precisely defined strain
rate $\dot{\epsilon}$ for long observation times \cite{Shenoy12042016,shenoy2015characterizing}. Briefly, the center-of-mass position of a target vesicle is trapped in real-time using fluorescence microscopy and model predictive control (MPC) algorithm. The MPC feedback controller determines the necessary flow rates required to achieve trapping at a specific point while maintaining a nearly constant strain rate in extensional flow and is achieved using computer-controlled pressure regulators. In this way, the Stokes trap can be used to confine vesicles under zero-flow conditions (with no external or net flow) or under non-zero net flow conditions \cite{Shenoy12042016,kumar2019orientation}, and the latter method was used to study non-equilibrium flow dynamics in the work.

\begin{figure}[t]
	\centering
	\includegraphics[width=0.4\textwidth]{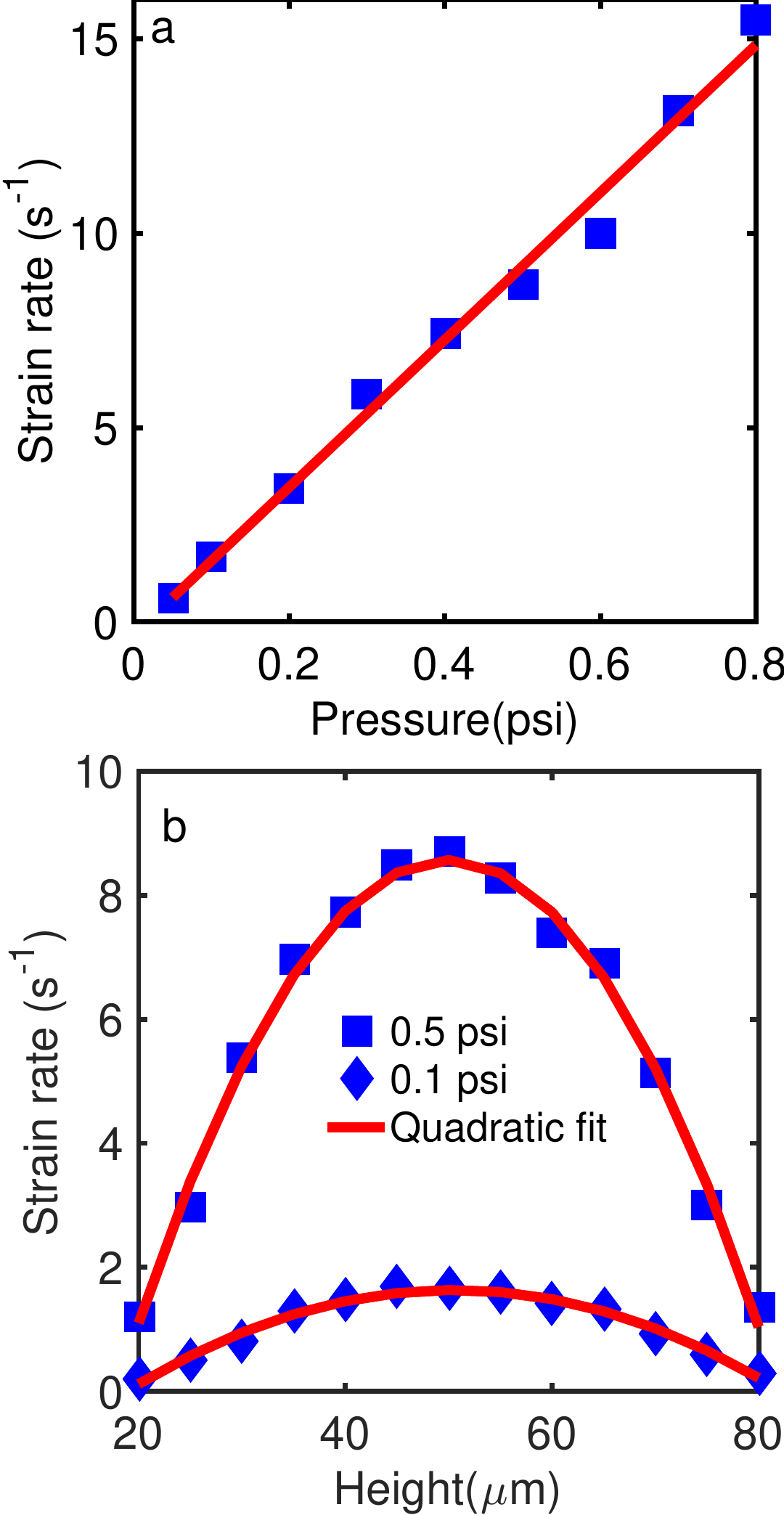}
	\caption{\label{vesicle_fig2}Flow-field characterization of cross-slot microfluidic devices. (a) Strain-rate determination at the center-plane of a cross-slot microfluidic device as a function of inlet pressure. Bead tracking experiments are performed in 100 mM sucrose buffer. (b) Strain-rate determination as a function of distance from the horizontal mid-plane in the device.}
\end{figure}

\subsection{Flow-field characterization}
Particle tracking velocimetry (PTV) is used to determine the fluid strain rates $\dot{\epsilon}$ as a function of the input pressure from the pressure regulators (Elveflow OB1-MkIII). Experimental characterization of the fluid strain rate is performed to ensure that the flow field is uniform in the vicinity of the stagnation point and enables determination of the capillary number $Ca=\mu_{out}\dot{\epsilon}R^3/\kappa_b$. A trace amount of fluorescent microbeads (2.2 $\mu$m diameter, Spherotech, 0.01\% v/v) was added to 105 mM sucrose buffer solutions ($\mu$ = 1.1 mPa-s, matched to the solution used for vesicle dynamics experiments) to enable particle tracking. Microfluidic devices are mounted on the stage of an inverted fluorescence microscope (Olympus IX71), which allows for real-time imaging of fluorescent beads using a high numerical aperture (1.45 NA, 63x) oil-immersion objective lens and a 100-W mercury arc lamp (USH102D, UShio). The sucrose buffer is introduced into microfluidic devices using the fluidic reservoirs, and images of bead positions are acquired using a CCD camera (GS3-U3-120S6M-C) as a function of the applied inlet pressure. For data shown in Fig. \ref{vesicle_fig2}a, the strain rate was determined at the center plane of the channel in the $z$-direction (direction orthogonal to the 2D flow plane). As shown in Fig. \ref{vesicle_fig2}a, the strain rate at the central plane of cross-slot device increases linearly with pressure over the characteristic range of strain rates used in this work. For data shown in Fig. \ref{vesicle_fig2}b, fluid strain rate was determined as a function of $z$-position by focusing through the depth of the device as a function of inlet pressure driving flow. A particle tracking and analysis program \cite{chen2014extending} is used to determine bead velocities for all trajectories, thereby enabling determination of the fluid strain rate $\dot{\epsilon}$ using a non-linear least square algorithm:
\begin{equation}
\begin{bmatrix} v_x\\v_y \end{bmatrix}=\begin{bmatrix}
\dot{\epsilon} & 0\\
0 & -\dot{\epsilon}
\end{bmatrix} \begin{bmatrix} x-x_0\\y-y_0 \end{bmatrix}
\label{strain_rate}
\end{equation}
where $v_x$, $v_y$, $x$, $y$ are velocities and positions in the $x$ and $y$ directions, respectively, and ($x_0, y_0$) is the location of the stagnation point in the 2D flow plane.  

\subsection{Bending modulus determination}
\subsubsection{Vesicle imaging in observation chamber}
For determination of bending modulus, vesicles are imaged in a secure-seal imaging spacer (Grace Bio-Labs, 7 mm diameter, 0.12 mm depth) using an inverted optical microscope (Olympus IX71) in epifluorescence mode equipped with a 63x oil immersion objective lens (NA 1.4, Zeiss Plan-Apochromat) and an electron multiplying charge coupled device (EMCCD) camera (Andor iXon-ultra, DU-897U-CSO, 512x512 pixel output). A 100-W mercury arc lamp (USH102D, UShio) was used as the excitation light source in conjunction with a neutral density filter (Olympus), a 530 $\pm$ 11 nm band-pass excitation filter (FF01-530/11-25, Semrock), and a 562-nm single-edge dichroic mirror (Di03-FF562-25 $\times$ 36, Semrock) in the illumination path. 

The vesicle suspension is first introduced into the spacer, and the top of spacer is then sealed with a coverslip to minimize evaporation and convection within the observation chamber. The temperature inside the chamber is measured using a thermocouple and found to be 22$^{\circ}$C for all experiments. The effect of gravity influencing the vesicle shape is negligible because of the nearly equivalent concentration of sucrose in the interior and exterior of the vesicle, yielding symmetry across the bilayer membrane. Imaging is performed at the central plane of the spacer, and the center-of-mass of vesicles remains nearly constant during an observation time of 30-60 s. Images are acquired over at least 30 s (acquisition frame rate of 30 Hz), which is much larger than the relaxation time of the slowest decaying mode of the membrane \cite{PhysRevA.36.4371}. The approximate order-of-magnitude relaxation time for a typical lipid membrane vesicle of size $R$ = 10 $\mu$m is $\approx$200 ms \cite{PhysRevA.36.4371}, yielding a bending modulus of $10^{-19}$J in a suspending medium with viscosity of 1 mPa-s. In this way, long observation times ensure that the available configurational modes of vesicles are given sufficient time to relax. For these experiments, unilamellar and defect-free vesicles are selected, and the fluctuating vesicles in the spacer are spatially isolated from their neighbors. 

\subsubsection{Contour detection and determination of $\kappa_b$}
We use the method proposed by P{\'e}cr{\'e}aux \textit{et al.} \cite{pecreaux2004refined} to determine vesicle bending modulus. In this way, we follow a rigorous selection criteria outlined in prior work \cite{meleard2011advantages,dahl2016experimental} that provides an unbiased procedure for rejecting unsuitable vesicles that do not fluctuate according to an analytical fluctuation spectrum given by the Helfrich model \cite{helfrich1973elastic}. Vesicle contours are first detected in each image with high precision using a custom MATLAB program that relies on intensity gradient maxima values to locate the edges ( ESI{\dag}, Fig. S1). The detected coordinate positions of the vesicle membrane $(x_i,y_i)$ in each movie frame are transformed to polar coordinates $(r_i,\theta_i)$ and projected into Fourier modes as follows: 
\begin{equation}
r\left (\theta \right )=R\left (1+\sum_{n=1}^{\infty }a_{n}\cos {\left (n\theta \right )}+b_{n}\sin \left (n\theta \right )  \right ) 
\label{eqn2}
\end{equation}
where \textit{R} is the radius of contour in each frame defined as: 
\begin{equation}
R=\frac {1}{2\pi}\sum_{i=1}^{N}\left (\frac {r_{i}+r_{i+1}}{2} \right )\left (\theta_{i+1}-\theta_{i} \right )
\label{eqn3}
\end{equation}
The magnitude of Fourier amplitudes is calculated as $c_{n}=\sqrt{a_{n}^{2}+b_{n}^{2}}$, and the mean square amplitude of fluctuation modes around a base spherical shape is given by: 
\begin{equation}
\left \langle \left | u\left (q_x \right ) \right |^{2} \right \rangle=\frac{\pi \left \langle R \right \rangle^{3}}{2} \left (\langle c_{n}^{2} \rangle-\langle c_{n} \rangle ^{2} \right )
\label{eqn4}
\end{equation}
where $q_{x}=n / \left \langle R \right \rangle$ is the wavenumber and $\left \langle R \right \rangle$ is the mean radius of contours determined over all images in a fluctuation experiment on a single vesicle. For determining the bending modulus $\kappa_b$ of vesicles, the following steps are performed:

(1) For each vesicle contour, the mean square amplitude of fluctuations is calculated using Eq. \ref{eqn4}. For this analysis, the behavior over modes $\textit{n}=6-25$ is examined (see ESI\dag for details).

(2) A one-sample Kolmogorov-Smirnoff test is used to check the exponential distribution of modes. In brief, vesicles maintain a constant volume and surface area over the the timescale of these experiments, so the Fourier modes in Eq. \ref{eqn4} are expected to be exponentially distributed. For the modes that pass this test, the experimental mean square amplitude $\left \langle \left | u\left (q_x \right ) \right |^{2} \right \rangle$ is calculated. In this way, the objective function $F$ is optimized: 
\begin{equation}
F\equiv \sum_{n=6}^{n=25}\frac{ \left \langle \left | u\left (q_x \right ) \right |^{2} \right \rangle- \left \langle \left | u_H\left (q_x \right ) \right |^{2} \right \rangle }{\sigma_{\left \langle \left | u\left (q_x \right ) \right |^{2} \right \rangle}^{2}}
\label{eqn5}
\end{equation}
where $\sigma_{\left \langle \left | u\left (q_x \right ) \right |^{2} \right \rangle}^{2}$ is the measured standard deviation of the experimental amplitudes $\left \langle \left | u\left (q_x \right ) \right |^{2} \right \rangle$ according to a procedure discussed in ESI\dag, and $\left \langle \left | u_H\left (q_x \right ) \right |^{2} \right \rangle$ is the modified form of Helfrich's spectrum after incorporating the effect of the finite camera integration time \cite{helfrich1973elastic}.

(3) The quantity $\left \langle \left | u\left (q_x \right ) \right |^{2} \right \rangle$ versus $q_x$ is plotted and analyzed for each vesicle to generate the experimental fluctuation spectrum. In this way, a two-parameter fit is performed using the modified form of Helfrich's spectrum that accounts for the effect of finite integration time of camera: 
\begin{eqnarray}
\left \langle \left | u_H\left (q_x \right ) \right |^{2} \right \rangle=\frac{1}{\pi}\int_{-\infty}^{\infty}\frac{kT}{4\mu_{out} q_{\perp}}\tau_m \frac{\tau_m^2}{\tau^2}\left [\frac{\tau}{\tau_m}+\exp{\left (\frac{-\tau}{\tau_m} \right )-1} \right ]dq_y 
\label{eqn6}
\end{eqnarray}
where $\tau_{m}^{-1}=\frac{1}{4\mu_{out} q_{\perp}} \left (\sigma q_{\perp}^{2}+\kappa_{b}q_{\perp}^{4} \right )$ and $ q_{\perp}=\sqrt{q_{x}^{2}+q_{y}^{2}}$. In this way, we determine the bending modulus $\kappa_b$ and membrane tension $\sigma$ for each vesicle \cite{pecreaux2004refined}. In this fitting procedure, the smallest value of the quantity $\left \langle \left | u\left (q_x \right ) \right |^{2} \right \rangle$ is taken to be $10^{-22}$ $m^{3}$, limited by the spatial resolution of camera (1 pixel $\approx$ 200 nm), determined in a separate experiment by measuring the fluctuation amplitudes of a stationary rigid fluorescently labeled polystyrene bead in the focal plane of the microscope.

We further estimate the correlation coefficient of the two parameters of fit ($\kappa_b$ and $\sigma$) (see ESI\dag for details). If the correlation coefficient $corr(\sigma,\kappa_b)<-0.85$, the vesicle is rejected for analysis because the membrane is generally taken to be too tense to provide an accurate estimate of the bending modulus \cite{meleard2011advantages}. In determining bending modulus $\kappa_b$, we only consider the images in which the average contour length does not change by more than 5\% to ensure the constant surface area and volume. Finally, we only consider quasi-spherical vesicles in the fluctuation analysis for estimation of bending modulus. The vesicles used in non-equilibrium flow experiments are highly deflated (non-spherical), though we follow prior work in assuming that the ensemble-averaged bending modulus measured for quasi-spherical vesicles is representative of all vesicles in the sample \cite{meleard2011advantages, dahl2016experimental}.

\subsection{Flow experiments in extensional flow}
Following flow field characterization and determination of equilibrium bending modulus $\kappa_b$, we studied the non-equilibrium deformation of vesicles in extensional flow. Non-equilibrium flow experiments were conducted using fluorescence microscopy at 10x magnification using an inverted optical microscope (Olympus IX71) with mercury lamp as the illumination source (100-W mercury arc lamp USH102D, UShio). Images were captured using a CCD camera (Pointgrey GS3 23S6M USB3 CMOS) at a frame rate of 30 Hz with an exposure time of 10 ms. A dilute vesicle suspension in sucrose buffer was introduced into the PDMS microfluidic device via sample tubing (PEEK tubing 1/16'' OD x 0.02'' ID) connected to fluidic reservoirs (Fig. \ref{vesicle_fig1}). The four fluidic reservoirs are pressurized using pressure transducers to drive the fluid into the microfluidic chip. The fluid inside and outside the vesicle (105 mM sucrose buffer) are density matched, so there is no significant drift of vesicles in the orthogonal direction ($z$-direction) during the timescale of the experiment. Vesicles were introduced into the cross-slot device by flowing through inlet channels at extremely low velocities such that vesicles are negligibly deformed prior to flow experiments.

\begin{figure}[t]
	\centering
	\includegraphics[width=0.4\textwidth]{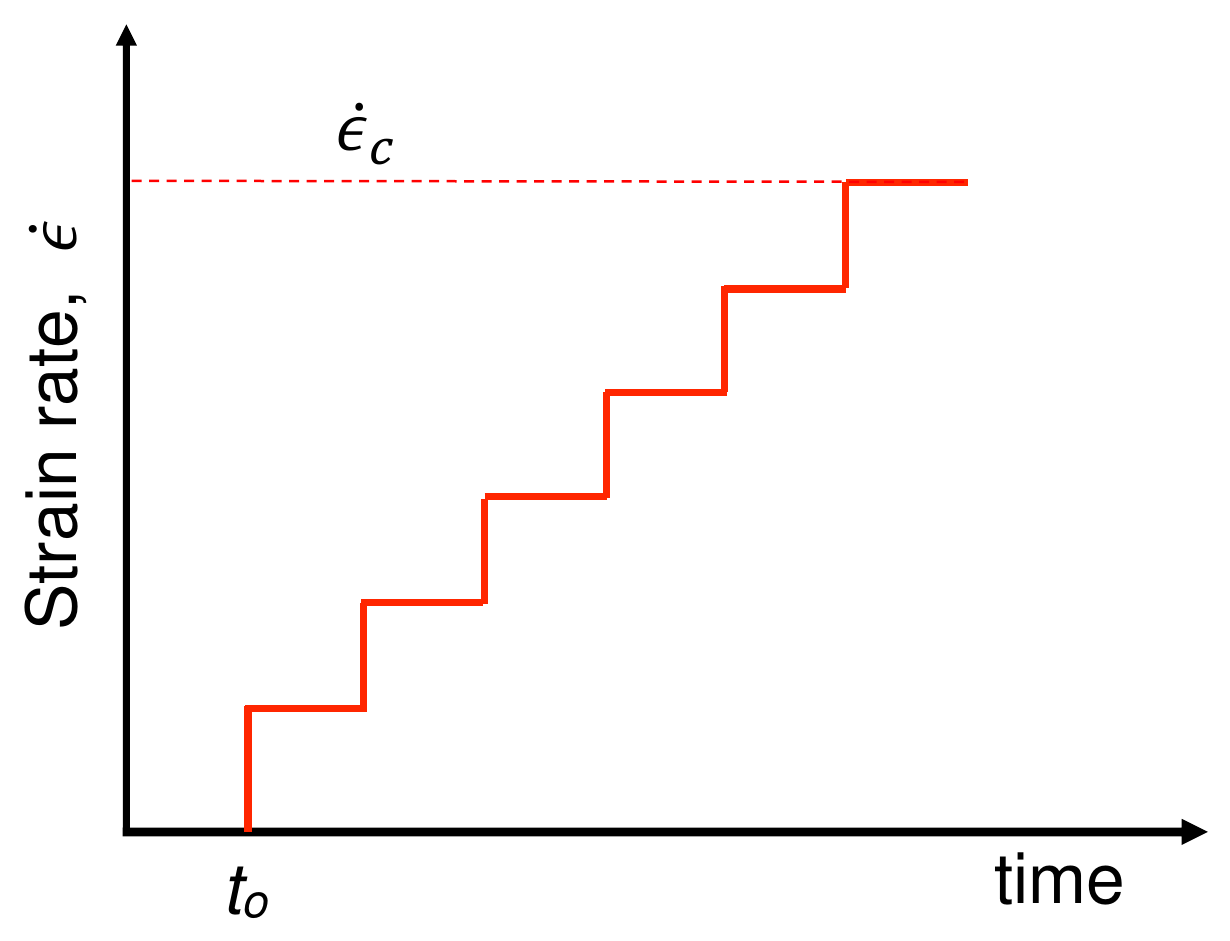}
	\caption{\label{vesicle_fig3} Flow deformation protocol and time-dependent strain rate schedule for the phase diagram experiments. Fluid strain rate is increased in a systematic step-wise fashion, and after each step change, vesicle shape is directly observed for $\approx$15-30 s. The critical strain rate $\dot{\epsilon}_{critical}$ is defined as the strain rate at which a vesicle undergoes a global shape transition.}
\end{figure}

In this work, we only consider vesicles that are unilamellar (via visual inspection of contour brightness and smoothness), defect-free, and completely isolated from neighboring vesicles. Multilamellar vesicles are observed in the sample, typically showing defects such as a daughter vesicle inside a parent vesicle, or lipid tubes protruding from the membrane, but these vesicles are not included in our analysis. Prior to performing a non-equilibrium flow experiment, individual vesicles are first trapped under zero-flow conditions for $\approx$ 15-30 s, thereby allowing the vesicle to relax for several seconds to ensure near-equilibrium behavior. During this step, the equivalent radius $R$ and reduced volume $\nu$ for each vesicle are measured under zero-flow conditions. Reduced volume $\nu$ is defined as the ratio of a vesicle's volume $V$ to the volume of an equivalent sphere with surface area $A=4\pi R^2$, such that:
\begin{equation}
\nu=\frac{3V\sqrt{4\pi}}{A^{3/2}}
\end{equation}
A reduced volume of $\nu$ = 1 corresponds to a perfectly spherical vesicle, whereas $\nu<1$ represents an osmotically deflated vesicle.

For these experiments, the membrane contour is located with a high precision using the edge detection method discussed in Section 2.4. To determine reduced volume, we follow the approach by Dahl \textit{et al.} \cite{dahl2016experimental}. In brief, the surface area and volume of a vesicle are estimated by revolution of the observed 2D membrane contour along the vesicle's short axis (ESI\dag, Fig. S2 and Fig. S3). The equilibrium shape of a vesicle is not always symmetric, so the volume and surface area are calculated from the top and bottom halves of the vesicle separately by numerical integration \cite{zhou2011stretching}, and the total surface area and volume are taken as the average value with uncertainty corresponding to one-half of the difference between the top and bottom halves. In this way, the equivalent vesicle radius $R$ and the reduced volume $\nu$ are determined from the mean of 100 images at equilibrium (ESI\dag, Fig. S4).

\begin{figure}[b]
	\centering
	\includegraphics[width=0.45\textwidth]{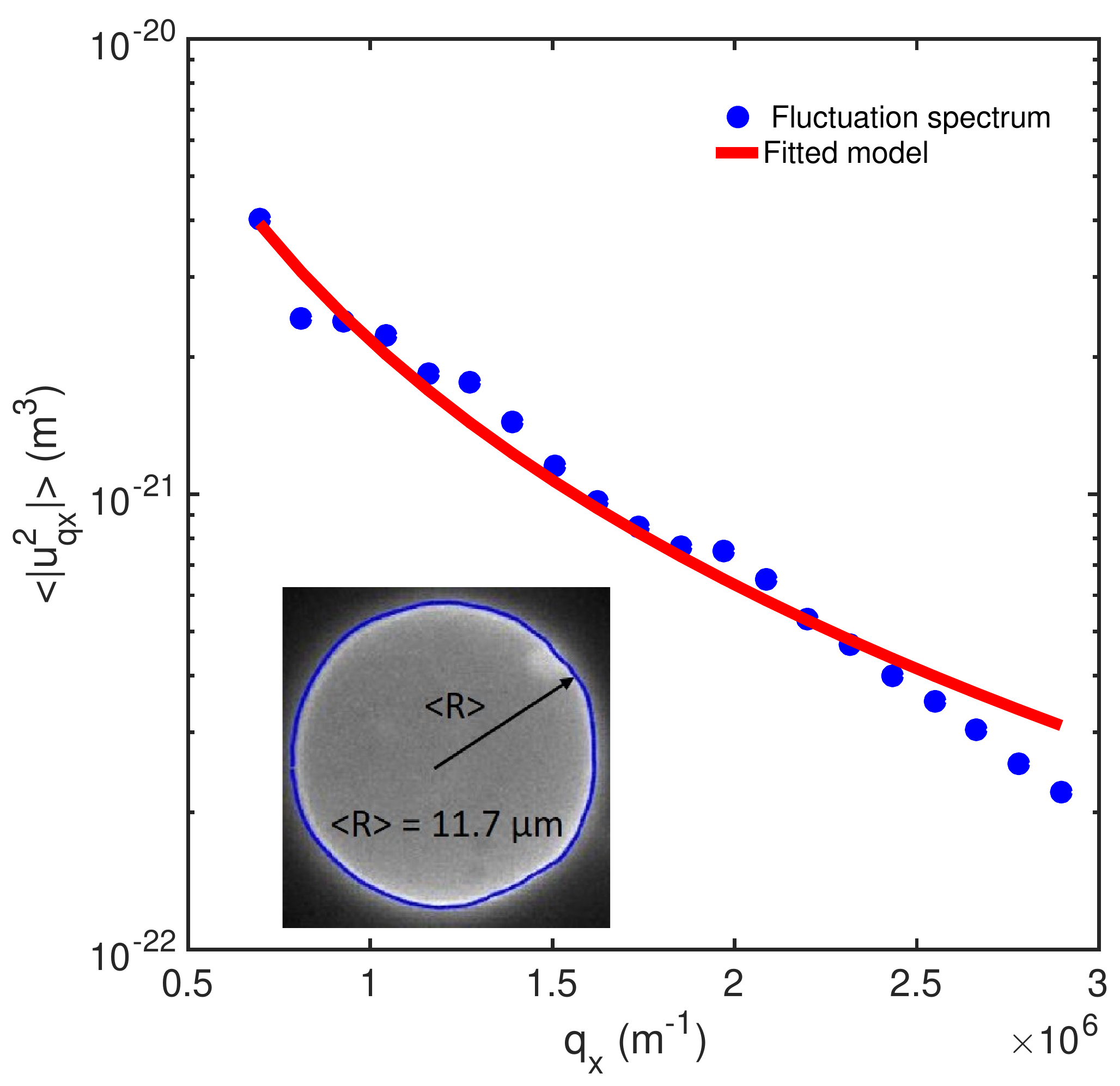}
	\caption{\label{vesicle_fig4} Analysis of membrane fluctuations for determining bending modulus $\kappa_b$. The amplitude of fluctuations $\left \langle \left | u\left (q_x \right ) \right |^{2} \right \rangle$ is plotted as a function of wave vector $q_x$ for a representative DOPC vesicle. The solid red line corresponds to the analytical model using Eq. 6, yielding $\kappa_b$ = 8.9 $\times$ 10$^{-20}$ J and membrane tension 3.9 $\times$ 10$^{-7}$ N/m (Inset): Detected contour of a fluctuating GUV at equilibrium using image processing methods.}
\end{figure}

Following determination of $R$ and $\nu$ for a single vesicle, the non-equilibrium flow experiment is performed by directly observing shape dynamics for the same individual vesicle in planar extensional flow. Vesicle dynamics in flow are governed by three dimensionless parameters: reduced volume $\nu$, capillary number $Ca$, and viscosity contrast $\lambda$. The capillary number $Ca$ is the ratio of the viscous forces to bending forces on the interface, such that:
\begin{equation}
Ca=\frac{\mu_{out}\dot{\epsilon}R^3}{\kappa_b} 
\end{equation}
where $\mu_{out}$ is exterior fluid viscosity, and the viscosity contrast $\lambda$ is the ratio of the fluid viscosities between the interior ($\mu_{in}$) and exterior ($\mu_{out}$) regions of a vesicle:
\begin{equation}
\lambda = \frac{\mu_{in}}{\mu_{out}}
\end{equation}

Using the Stokes trap, the fluid strain rate is increased in a systematic step-wise fashion (Fig. \ref{vesicle_fig3}) by changing the pressure difference $\delta P$ between the inlet and outlet channels in the microfluidic device (Fig. \ref{vesicle_fig1}b). After each step increase in the flow rate, a trapped vesicle is observed for $\approx$ 15-30 s and shape fluctuations are directly observed. The observation time at each constant strain rate is longer than the time required for the shape changes to occur, estimated from linear stability analysis \cite{zhao2013shape,narsimhan2015pearling}. In this way, we systematically study vesicle shape transitions across a wide range of parameters in $(\nu, Ca)$ space with high resolution between experimental data points along the $Ca$-axis in the flow-phase diagram. 

\section{Results and discussion}
\subsection{Bending modulus estimation}

\begin{figure*}[t]
	\centering
	\includegraphics[width=1.0\textwidth]{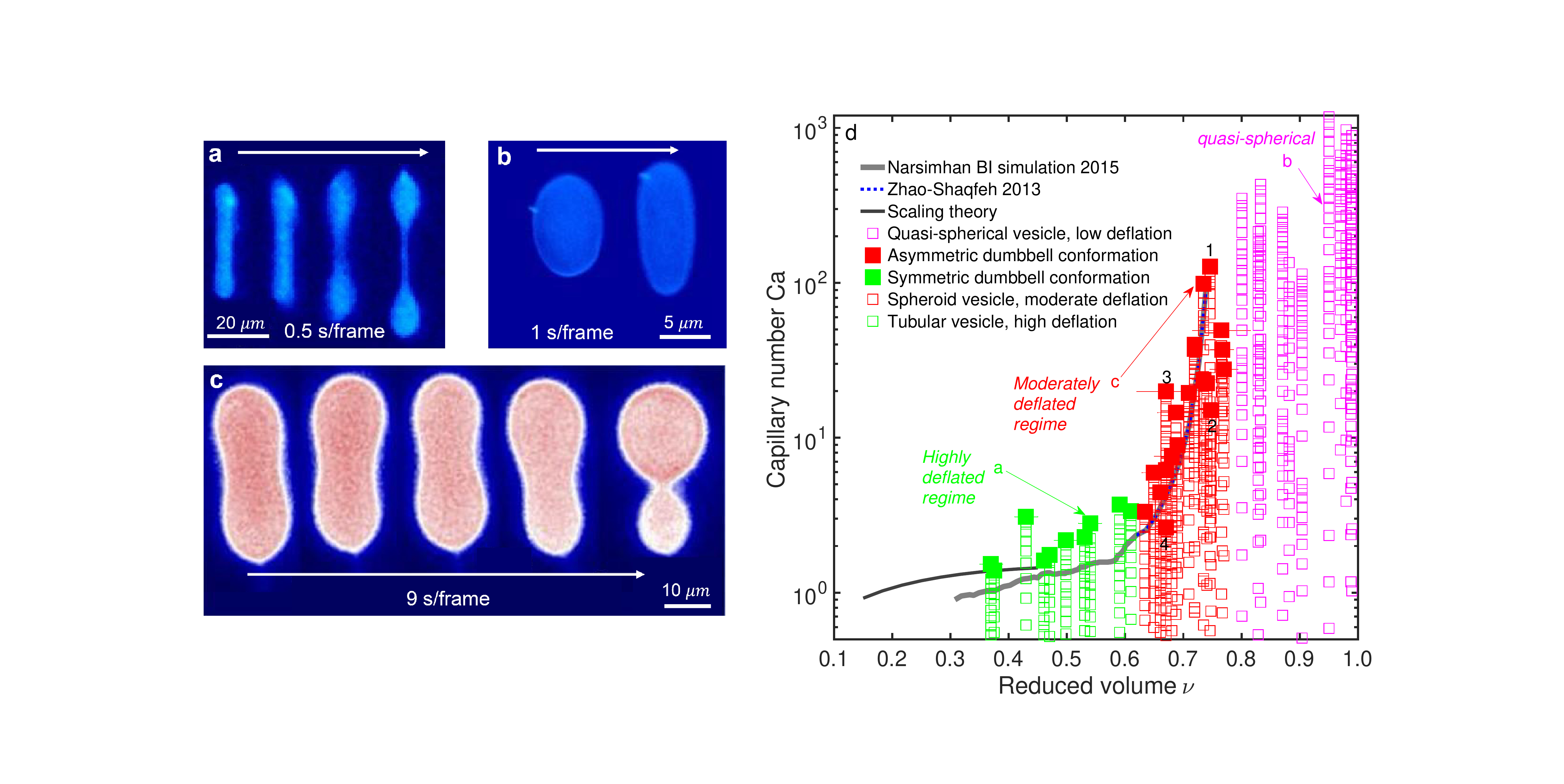}
	\caption{\label{vesicle_fig5} Flow-phase diagram for vesicle dynamics in extensional flow. Time series of images showing vesicle shape changes in extensional flow for: (a) a vesicle in the highly deflated regime $\nu=0.53$, having a tubular shape at equilibrium and undergoing a symmetric dumbbell shape transition at $Ca=2.3$, (b) a vesicle in the weakly deflated regime $\nu=0.95$, having a quasi-spherical shape at equilibrium and maintaining a stable ellipsoid shape upon extension up to $Ca$ of $\approx$1000, and (c) a vesicle in the moderately deflated regime $\nu=0.73$, having a spheroid shape at equilibrium and undergoing an asymmetric dumbbell shape transition at $Ca=98.7$. (d) Flow-phase diagram of vesicles in planar extensional flow as a function of reduced volume $\nu$ and capillary number $Ca$ at a viscosity ratio $\lambda$ = 1. Open green squares represent vesicles in the highly deflated regime $\nu<0.60$ (tubular shape at equilibrium) for $Ca<Ca_C$ at which a vesicle does not undergo shape instability. Filled green squares represent the $Ca_c$ phase boundary at which a tubular to symmetric dumbbell transition occurs. Open red squares represent vesicles in the moderately deflated regime $0.60<\nu<0.75$ (spheroid shape at equilibrium) for $Ca<Ca_C$ at which a vesicle does not undergo a shape instability. Filled red squares represent the $Ca_c$ phase boundary at which a spheroid to asymmetric dumbbell transition occurs. Open magenta squares represent vesicles in the weakly deflated regime $\nu>0.75$ where vesicles have quasi-spherical shape at equilibrium and transition to a stable ellipsoid shape. The grey curve represents the phase boundary from boundary integral simulations \cite{narsimhan2014mechanism}.}
\end{figure*}

We began by determining the average bending modulus for an ensemble of DOPC lipid vesicles using the procedure described in the Experimental Methods (Section 2.4). In brief, this method relies on analyzing membrane fluctuations for weakly deflated vesicles at equilibrium (no flow conditions), followed determination of bending modulus $\kappa_b$ and membrane tension $\sigma$ using a two-parameter fit to the Helfrich model given by Eq. \ref{eqn6}. The amplitude of membrane thermal fluctuations $\left \langle \left | u\left (q_x \right ) \right |^{2} \right \rangle$ as a function of wave vector $q_x$ is shown for a characteristic lipid vesicle in Fig.\ref{vesicle_fig4}. Using this approach, we determined an average bending modulus of $\kappa_b$ = $\left( 9.17 \pm 0.20 \right) \times 10^{-20}$ J ($N$ = 21). The average value of membrane tension was found to be $\sigma$ = $\left( 1.9 \pm 0.20 \right) \times 10^{-7}$ N/m ($N$ = 21), which is consistent with prior work reported in literature \cite{dahl2016experimental}.

The experimentally determined value of the bending modulus for DOPC vesicles ($\kappa_b$ = $9.17 \times 10^{-20}$ J, DOPC with 0.12 mol\% DOPE-Rh, 100 mM sucrose, $T$ = 24$^\circ$C) is in reasonable agreement with the bending modulus measured for pure DOPC vesicles ($\kappa_b$ = $ 9.1 \times 10^{-20}$ J, 300 mM sucrose/307 mM glucose, $T$ = 25$^\circ$C) by Zhou \textit{et al.} \cite{zhou2011stretching}, which suggests that the bending modulus for DOPC vesicles does not significantly depend on sugar concentration over the relatively narrow the range of 100-300 mM. Indeed, low angle X-ray scattering measurements by Nagle \textit{et al.} \cite{nagle2016sugar,nagle2015true} have recently shown that the bending modulus of DOPC vesicles does not depend on sucrose concentration in the range between 100-450 mM. Gracia \textit{et al.} measured the bending modulus of pure DOPC vesicles (10 mM glucose, $T$ = 25$^\circ$C) to be $\kappa_b$ = 10.8 $\times 10^{-20}$ J, which is consistent with the value of $\kappa_b$ measured in this work at a higher sucrose concentration of 100 mM. Prior work \cite{shchelokovskyy2011effect,vitkova2006sugars} has shown that increasing the sugar concentration from 10 mM to 100 mM decreases the value of bending modulus by a factor of two, though our results tend to show less deviation in $\kappa_b$ over this range of sucrose concentration. Indeed, such variabilities in experimental measurements of bending moduli for DOPC vesicles have been reported in prior work \cite{nagle2013introductory}.

The DOPC vesicles in this work contain an exceedingly small amount of fluorescently labeled lipid (0.12 mol\% DOPE-Rh), which suggests that such a low concentration of labeled lipid does not substantially alter the bending modulus of the membrane compared to pure DOPC vesicles \cite{bouvrais2010impact}. Our method for determining bending modulus relies on a fairly strict set of statistical rejection criteria for excluding vesicles that do not conform to an analytical model (Eq. \ref{eqn6}), which yields a relatively narrow distribution in bending moduli values across the ensemble. Nevertheless, variability in bending modulus between individual vesicles can be attributed to light-induced peroxide formation in GUVs and/or precision of membrane edge detection in vesicle images \cite{bouvrais2010impact}. Broadly speaking, the experimentally measured values of $\kappa_b$ in this work are consistent with prior work reported for DOPC vesicles \cite{nagle2016sugar,dimova2014recent,gracia2010effect}. 

\begin{figure*}[t]
	\centering
	\includegraphics[width=1.0\textwidth]{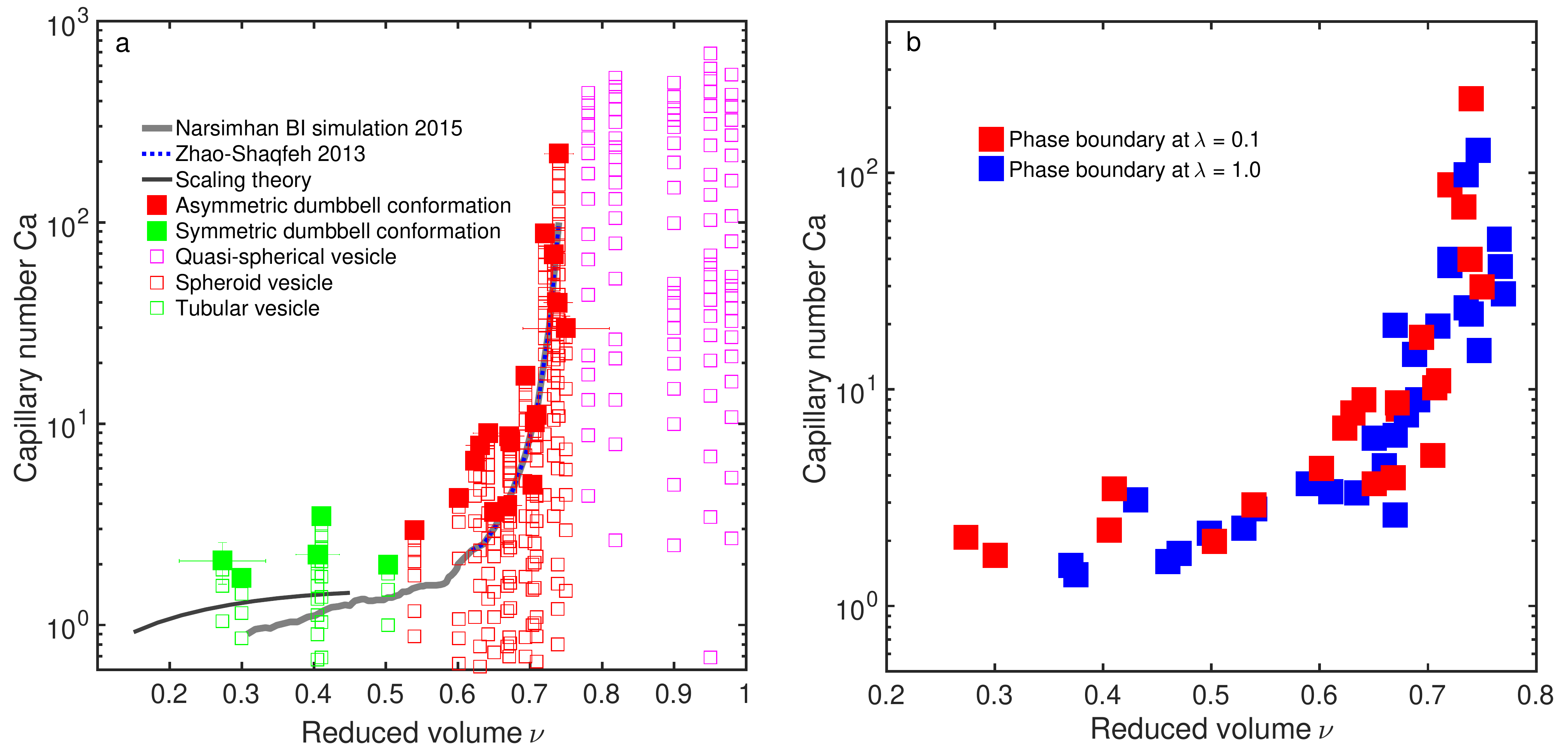}
	\caption{\label{vesicle_fig6} Flow-phase diagram for vesicle dynamics in extensional flow at viscosity ratio $\lambda=0.1$. (a) Flow-phase diagram of vesicle conformations as a function of reduced volume $\nu$ and capillary number $Ca$ at viscosity ratio $\lambda=0.1$. Green, red and magenta markers have the same meaning as Fig. \ref{vesicle_fig5}. (b) Comparison between phase boundaries for viscosity ratios $\lambda=1$ and $\lambda=0.1$.}
\end{figure*}

\subsection{Non-equilibrium flow-phase diagrams}
Following determination of bending modulus $\kappa_b$, we further studied the non-equilibrium dynamics and conformation phase transitions of vesicles in extensional flow over a wide range of reduced volume $\nu$ and capillary number $Ca$ for a uniform viscosity contrast $\lambda$ = 1 (Fig. \ref{vesicle_fig5}). Using the Stokes trap, we confined single vesicles near the stagnation point of planar extensional flow and observed the non-equilibrium shape dynamics while systematically increasing the strain rate $\dot{\epsilon}$ in a scheduled fashion (Fig. \ref{vesicle_fig3}). In this way, vesicles were observed to adopt a wide variety of shapes in flow, including a symmetric dumbbell shape (Fig. \ref{vesicle_fig5}a, Movie S1 ESI\dag), an asymmetric dumbbell shape (Fig. \ref{vesicle_fig5}c, Movie S2 ESI\dag), and a stable ellipsoidal shape (Fig. \ref{vesicle_fig5}b, Movie S3 ESI\dag) depending on the flow strength $Ca$ and amount of membrane floppiness $\nu$. 

Fig. \ref{vesicle_fig5}a shows a characteristic time series of images for a highly deflated ($\nu$ = 0.53) vesicle initially in a tubular shape under zero flow conditions. In the presence of extensional flow, the vesicle stretches along the extensional axis and eventually transitions to a symmetric dumbbell shape at $Ca$ = 2.3. Once the shape change occurs, the vesicle is observed to reach a steady-state conformation in flow. Similarly, Fig. \ref{vesicle_fig5}c shows a characteristic time series of images for a moderately deflated ($\nu$ = 0.73) vesicle initially in a spheroidal shape, eventually transiting to an asymmetric dumbbell shape at $Ca$ = 98.7. Finally, Fig. \ref{vesicle_fig5}b shows a time series of images for a quasi-spherical vesicle that largely retains an ellipsoidal shape as $Ca$ increases and does not undergo a transition into a dumbbell shape.

The experimental flow-phase diagram for vesicle shapes in extensional flow is shown in Fig. \ref{vesicle_fig5}d. Our results reveal three distinct dynamical regimes in the ($\nu$, $Ca$) plane attained by lipid vesicles. In general, highly deflated ($\nu<0.60$) and moderately deflated ($0.60<\nu<0.75$) vesicles are observed to transition into symmetric or asymmetric dumbbell shapes, respectively, at a critical strain rate $\dot{\epsilon_c}$ (Movie S1,S3 ESI\dag). The critical capillary number $Ca_c$ for the vesicle shape transition depends on the reduced volume $\nu$. As shown in Fig. \ref{vesicle_fig5}d, the filled green symbols (red symbols) represent the symmetric (asymmetric) dumbbell shape transition for vesicles with reduced volume $\nu<0.60$ ($0.60<\nu<0.75$). The vertical set of open green and red squares represent data obtained by systematically stepping strain rate using the Stokes trap for $Ca$ values below the critical value for a shape transition. At higher reduced volumes ($\nu>0.75$), vesicles retain a stable ellipsoidal shape regardless of $Ca$ and do not undergo a symmetric/asymmetric dumbbell shape change over the entire range of $Ca$. The grey curve shows the predicted stability boundary from boundary-integral simulations \cite{narsimhan2014mechanism,zhao2013shape,narsimhan2015pearling}, which is in good agreement with our experimental data.

In general, the flow-phase diagram reveals three distinct regimes in vesicle shape dynamics defined by reduced volume $\nu$: (i) $\nu<0.60$, (ii) $0.60<\nu<0.75$, and (iii) $\nu>0.75$ corresponding to transitions to a symmetric dumbbell, asymmetric dumbbell, or stable ellipsoid shape, respectively. Interestingly, the critical capillary number $Ca_c$ required to trigger a shape transition decreases with higher levels of deflation (decreasing $\nu$). These observations are consistent with prior experimental work \cite{kantsler2008critical,dahl2016experimental} and numerical simulations on vesicle dynamics in extensional flow \cite{narsimhan2015pearling,narsimhan2014mechanism}. Moreover, the predicted phase boundary from a scaling analysis in prior work \cite{narsimhan2015pearling} is also shown in Fig.\ref{vesicle_fig5}d, which appears to be in qualitative agreement with experiments. 

Our experimental results also reveal some degree of variability in the behavior of vesicle shape transitions near the critical stability boundary. For example, vesicles marked as `\textbf{1}' and `\textbf{2}' in Fig. \ref{vesicle_fig5}d have approximately the same reduced volume $\nu\approx0.74$, but they transition to an asymmetric dumbbell shape at different values ($Ca$=133 and $Ca$=17.8, respectively). Similarly, vesicles marked as `\textbf{3}' and `\textbf{4}' undergo an asymmetric shape transition at $Ca$ numbers slightly above and below the curve predicted from simulations. In general, such variability in vesicle dynamics near the phase boundary can arise due to several reasons. First, $Ca$ is defined based on an ensemble averaged value of bending modulus $\kappa_b$ determined from thermal fluctuation analysis of quasi-spherical vesicles at equilibrium. In the non-equilibrium flow experiments, vesicles are osmotically deflated and have non-spherical shapes, which may result in differences in bending modulus on an individual vesicle basis. Moreover, our method of estimating reduced volume $\nu$ by assuming a 2D contour for vesicles as a body of revolution generally ignores thermal wrinkles in the vertical direction ($z$-direction), which may introduce minor variability in determining $\nu$ \cite{deschamps2009phase,zhou2011stretching}. Finally, numerical simulations of vesicle shape dynamics do not include thermal fluctuations of the vesicle membrane, which may lead to differences between experimental results and numerical predictions. Indeed, our results show that the role of thermal fluctuations may be important in describing the nature of vesicle shape transitions in flow (Movies S1, S3, see ESI\dag for details). 

To investigate the influence of viscosity ratio $\lambda$ on the stability boundary, we performed an additional set of experiments by increasing the viscosity of the suspending medium by adding glycerol, such that the viscosity ratio $\lambda$ = 0.1. Fig. \ref{vesicle_fig6}a shows the flow-phase diagram for DOPC vesicles in extensional flow as a function of $Ca$ and $\nu$ at $\lambda$ = 0.1. Overall, the dynamic behavior of vesicles at $\lambda$ = 0.1 was similar to that observed at $\lambda$ = 1.0. To quantitatively compare the dynamic behavior of vesicles at different viscosity ratios, we plotted the stability boundary for $\lambda$ = 1.0 and 0.1 in Fig. \ref{vesicle_fig6}b. The difference between these curves is not statistically significant as determined by a Mann-Whitney test ($p>0.05$). Overall, these results suggest that the onset of the symmetric and asymmetric dumbbell instabilities is independent of the viscosity ratio due to membrane area incompressibility. This can be understood by the fact that the flow in the base state interior to the vesicle is generally not sensitive to the viscosity ratio because the interface is immobile due to a constant membrane area. It should be noted that the dynamic behavior of vesicles with a molecularly thin membrane is markedly different compared to immiscible drops with a simple liquid-liquid interface. In the case of immiscible drops, the viscosity ratio plays a key role in their dynamics, such that the critical capillary number required for drop burst instability is a strong function of the viscosity ratio $\lambda$ \cite{bentley1986experimental}.

\section{Conclusions}
In this work, we experimentally determine the flow-phase diagrams for vesicles in extensional flow with high resolution in $(Ca, \nu)$ space using a Stokes trap. Our results show that vesicles undergo symmetric and asymmetric dumbbell shape transitions depending on $Ca$ and $\nu$ over a wide range of conditions. Quantitative characterization of the phase diagram reveals three distinct dynamical regimes for vesicles in extensional flow namely, a tubular to symmetric dumbbell transition, a spheroid to asymmetric dumbbell transition, and quasi-spherical to stable ellipsoid depending on the value of reduced volume. We further demonstrate that the phase boundary for shape transitions in flow is insensitive to viscosity contrast between vesicle interior and exterior. Due to the presence of the incompressible molecularly thin lipid bilayer membrane, vesicles exhibit very different dynamics compared to liquid drops in flow.

Importantly, the trapping method used in this work allows vesicles to reach a steady-state conformation in extensional flow after experiencing a global change in shape. We emphasize that such experimental precision was enabled by using the Stokes trap, which allows for the long-time observation of single or multiple particles in an externally imposed flow. An intriguing question relates to vesicle dynamics at flow rates exceeding the critical capillary number $Ca_c$. Upon increasing the flow rates above $Ca_c$, we anticipate that vesicles will continue to stretch and will likely undergo large deformations to extremely high large aspect ratios (ratio of a vesicle's stretched length along the extensional axis to the equilibrium length). In future work, it will be interesting to investigate if additional membrane-bound soft materials such as polymersomes (polymer vesicles), capsules, or cells undergo similar shape changes under flow. Overall, our work establishes the utility of Stokes trap as a tool for investigating vesicle dynamics and opens new avenues for investigating the non-equilibrium dynamics of soft deformable particles in strong flows.  

\section*{Acknowledgements}
We thank Vivek Narsimhan for insightful discussions and Anish Shenoy and Kejia Chen for help in implementing the Stokes trap and analysis of vesicle fluctuations. This work was funded by National Science Foundation (NSF) through grant CBET PMP 1704668.
		
	%\bibliography{refs}

%apsrev4-2.bst 2019-01-14 (MD) hand-edited version of apsrev4-1.bst
%Control: key (0)
%Control: author (8) initials jnrlst
%Control: editor formatted (1) identically to author
%Control: production of article title (0) allowed
%Control: page (0) single
%Control: year (1) truncated
%Control: production of eprint (0) enabled
%

\end{document}